\documentclass[12pt]{article}
\usepackage{amsmath,amssymb}
\newcommand{\bq}{\begin{eqnarray}}
\newcommand{\eq}{\end{eqnarray}}
\newcommand{\be}{\begin{equation}}
\newcommand{\ee}{\end{equation}}
\newcommand{\ra}{\rightarrow}

\newcommand{\ov}{\overline}
\newcommand{\nd}{{\ov N}_c}
\newcommand{\la}{ \Lambda_Q }

\newcommand{\te}{d^2\theta}
\newcommand{\ote}{d^2\ov \theta}

\newcommand{\cm}{{\cal M}_{\rm ch}}
\newcommand{\cma}{{\cal M}^{\it l}_{\rm ch}}
\newcommand{\cmb}{{\cal M}^{\rm h}_{\rm ch}}
\newcommand{\bo}{\rm{b_o}}

\newcommand{\lym}{\Lambda_{YM}}
\newcommand{\ql}{Q^{\it l}}
\newcommand{\bl}{\it\bar l}
\newcommand{\oql}{{\ov Q}_{\bl}}
\newcommand{\qh}{Q^{\rm h}}
\newcommand{\bh}{\rm\bar h}
\newcommand{\oqh}{{\ov Q}_{\bh}}
\newcommand{\mh}{m_{\it h}}
\newcommand{\mhp}{m_{\it h}^{\rm pole}}
\newcommand{\hmh}{{\hat m}_{\it h}}
\newcommand{\ml}{m_{\it l}}
\newcommand{\hml}{{\hat m}_{\it l}}

\newcommand{\nl}{N_{\it l}}
\newcommand{\nh}{N_{\rm h}}

\newcommand{\hcma}{\hat{\cal M}_{\rm ch}^{\it l}}
\newcommand{\hcmb}{\hat{\cal M}_{\rm ch}^{\it h}}
\begin{document}

\begin{center}
{\bf On Mass Spectrum in SQCD. \, Unequal quark masses}
\end{center}
\vspace{1cm}
\begin{center}\bf Victor L. Chernyak \end{center}
\begin{center}(e-mail: v.l.chernyak@inp.nsk.su) \end{center}
\begin{center} Budker Institute of Nuclear Physics, 630090 Novosibirsk, Russia
\end{center}
\vspace{1cm}

\begin{center}Abstract \end{center}

${\cal N}=1$ SQCD with $N_c$ colors and two types of light quarks, $\nl$ flavors with
smaller masses $\ml$ and $\nh=N_F-\nl$ flavors with larger masses $\mh$\,, $N_c<N_F<3N_c$
\,, \,$0<\ml\leq\mh\ll\la$\,, is considered within the dynamical scenario in which quarks
can form the coherent colorless diquark-condensate. There are several phase states at
different values of parameters $r=\ml/\mh\,,\,\nl$\,, and $N_F$.
Properties of these phases and the mass spectra therein are described.\\

\vspace{2cm}
\rm

{\bf 1\,. Introduction}.\\

In the present paper  the results obtained in \cite{ch} for equal quark masses are
generalized to the case of unequal masses. We do not consider here the most general
case of arbitrary quark masses. Only one specific (but sufficiently representative) choice of
unequal masses is considered\,: there are $\nl\neq N_c$ flavors with the smaller masses $m_l$
and $\nh=N_F-\nl$ flavors with the larger masses $\mh\geq\ml>0$\,,\,\,  $N_c<N_F<3N_c$\,. Some
abbreviations used below\,: DC means the diquark-condensate, HQ stands for a heavy quark,
the $\it l$-quarks means the quarks with the smaller masses $\ml$\,,\,the $\rm h$-quarks
are those with the larger masses $\mh$\,. The masses $\ml$ and $\mh$ are the running
current quark masses normalized at $\mu=\la$, and $\cma$ or $\cmb$ are the chiral diquark
condensates of the $\it l$ or $\rm h$-quarks also normalized at $\mu=\la$,\,\, $\la$
(independent of quark masses) is the scale parameter of the gauge coupling constant. All quark masses
are small, $0<\ml\leq \mh\ll \la$.

So, the whole theory can be considered as being defined by three numbers $N_c\,,\,\,N_F\,,\,
\,\nl$ and by three dimensional parameters $\la\,,\,\,\ml\,,\,\,\mh$ (i.e. all dimensional
observables  will be expressed through these three).\\

As will be shown below, within the dynamical scenario used, there are different phase states
in this theory at different values of parameters $r=\ml/\mh\leq 1\,,\,\,\nl$\,, and $N_F$\,:\\
a)\,\,the  $\rm DC_{\it l}-DC_{\rm h}$ phase appears at $\mh^{\rm pole}\ll\cmb<\cma\ll\la\,$\,
in both cases $\nl>N_c$ and $\nl<N_c$\,\,($\mh^{\rm pole}$ is the perturbative pole mass of
the $\rm h$-quarks)\,;\\
b) \,\,the $\rm DC_{\it l}-HQ_{\rm h}$ phase appears at  $\cmb\ll\cma\ll\mh^{\rm pole}\ll
\la$\,, and at $\nl>N_c$ only\,;\\
c) \,\,the another regime of the $\rm DC_{\it l}-HQ_{\rm h}$ phase appears  at $\cmb\ll
\mh^{\rm pole}\ll\cma\ll \la$\,, and in both cases $\nl>N_c$ and $\nl<N_c$\,;\\
d) \,\,the ${\rm Higgs_{\it l}-DC_{\rm h}}$ or ${\rm Higgs_{\it l}-HQ_{\rm h}}$\,-\,phases
appear at $\cma\gg\la$\,, and at $\nl<N_c$ only\,.\\

It is implied that a reader is familiar with the previous paper \cite{ch}, because all the
results from \cite{ch} are used essentially in this paper.\\

The paper is organized as follows. The properties of the $\rm DC_{\it l}-DC_{\rm h}$
phase are considered in section 2. The phase $\rm DC_{\it l}-HQ_{\rm h}$ (in two regimes) is
considered in sections 3 and 4. The phases ${\rm Higgs_{\it l}-DC_{\rm h}}$ and ${\rm Higgs_
{\it l}-HQ_{\rm h}}$ with higgsed $\it l$-quarks are considered in section 5. The section
6 contains a short conclusion.\\

{\bf 2\,. \,\,The \,\boldmath{$\rm DC_{\it l}-DC_{\rm h}$} phase}.\\

Let us recall first the effective Lagrangian for equal mass quarks and just below the
physical threshold at $\mu <\mu_H=\cm$, after the evolution of all quark degrees of freedom has been 
finished  \cite{ch}\, ($\bo=3N_c-N_F,\, \nd=N_F-N_c$\,, see also the footnote 5 in \cite{ch})\,:
\bq
L=\int \te \,\ote \, \Biggl \{ \rm {Tr}\,\sqrt {\Pi^\dagger \Pi}+Z_Q\,{\rm Tr}\Biggl ( Q
^\dagger e^{V} Q+ {\ov Q}^\dagger e^{-V} {\ov Q}\Biggr ) \Biggr \}+
\int \te \,\Biggl \{ -\frac{2\pi}{\alpha(\mu)}\,S +W_Q \Biggr \} \,, \nonumber
\eq
\bq
W_Q=\Biggl ( \frac{\rm {\det}\,\Pi}{\la^{\bo}}\Biggr )^{1/\ov N_c}\Biggl \{ \rm {Tr} \Bigl
(\ov Q \,\Pi^{-1}\, Q\Bigr )-N_F \Biggr \}+\rm {Tr}\Bigl (m_Q \Pi \Bigr )\,, \quad Z_Q=
\frac{m_Q}{\cm}\,.
\eq
Here\,: $(m_Q)_i^{\ov j}\equiv m_Q(\mu=\la)_i^{\ov j}$, where $m_Q(\mu)_i^{\ov j}$ are the running
quark masses, and $\langle \Pi^i_{\ov j}\rangle=\langle ({\ov Q}_{\ov j}Q^i)_{\mu=\la} \rangle
\equiv (\cm^2)^i_{\ov j}$. For equal quark masses\,: $(m_Q)_i^{\ov j}=m_Q\,\delta_i^{\ov j},\,\,
(\cm^2)^i_{\ov j}=\cm^2\,\delta^i_{\ov j}\,,\,\, m_Q\cm^2=\langle S\rangle=(\la^{\bo}\det
m_Q)^{1/N_c}$\,; (here and everywhere below for DC -phases\,: as for concrete forms of the pion
Kahler terms , see the footnote 5 in \cite{ch}).

Well above the highest physical threshold, $\mu_H\ll\mu\ll \la $\,, the quark fields ${\ov Q},\,Q$
describe the original quarks with the small running current masses $m_Q(\mu)$, while  below the
threshold they become the fields of heavy quarks with the large constituent masses $\mu_C=\cm$.
\footnote{\,
The Konishi anomaly \cite{Konishi} for the canonically normalized constituent quark fields
$C=Q/Z_Q^{1/2}$ and ${\ov C}={\ov Q}/Z_Q^{1/2}$ looks as\,: $\langle{\ov C}C\rangle=\langle
S\rangle /\mu_C$. But the form of its explicit realization is a matter of convention. One
convention is that it is realized directly through the one-loop triangle diagram with the
heavy constituent quarks forming a loop and emitting two gluinos. Second convention is that the
one loop constituent quark contributions into the vacuum polarization are transferred to the
gluon kinetic term at a first stage, and then there appears a term $ S\ln {\mu_C}$, while the
quark term in $W_Q$ in (1) has to be used now for calculations with the valence constituent
quarks only. The Konishi anomaly originates then from this vacuum polarization term and looks
now as\,: $\langle {\ov C}C \rangle=\langle\partial /\partial {\mu_C}\,(S\ln {\mu_C)}
\rangle=\langle S \rangle /{\mu_C}$.
}

The fields $\Pi$ are defined as "the light part of ${\ov Q}Q$". I.e. well above the threshold, when
the large constituent mass of quarks is not yet formed, $\Pi$ and ${\ov Q}Q$ are both the same
living diquark operator of light quarks, so that ${\ov Q}\Pi^{-1}Q$ is a unit c\,-\,number
matrix, and the projector ${\cal P}=\rm Tr (\ov Q \,\Pi^{-1}\, Q\Bigr )-N_F=0$. Moreover, the
term $(\det \Pi/\la^{\bo})^{1/\nd}$ is dominated by contributions of light quantum quark fields,
and represents not a constant mass, but a living interaction. But below the threshold, at
$\mu < \cm$, after the large constituent mass $\cm$ appeared, the light $\Pi$ and heavy ${\ov
Q}Q$ become quite different, so that ${\cal P}$  becomes a non-trivial nonzero term. Besides,
below the threshold all $N_F^2$ fields $\Pi$ become "frozen", in a sense that all of them
contain the large c-number vacuum part $\cm^2$ and the light quantum pion fields $\pi$ with the
small masses $m_Q$, whose contributions to amplitudes are smaller, $|\pi|\leq\mu <\cm$. As a
result, the whole term $(\det \Pi/\la^{\bo})^{1/\nd}(Z_Q\Pi)^{-1}$ in $W_Q$ is dominated now by
the c-number vacuum part which becomes the large constituent mass $\cm$ of the quark fields
$Q$ and $\ov Q$.\\

Let us start with $\ml=\mh$ and begin to make $\mh > \ml$, so that a gap appears between
$\cma > \cmb$\,:\,{\footnote{\,
But, to remain in the same $\rm DC$ - phase for all flavors, there will be a restriction on
the values of  $\ml$ and $\mh$, such that $r=\ml/\mh$ can't be too small. The
explicit form of this restriction will be presented below.
}
\bq
\Bigl (\cma\Bigr )^2=\frac{1}{\ml}\Bigl (\la^{\bo}\det m\Bigr )^{1/N_c}=\la^{\bo/N_c}\,
\ml^{(\nl-N_c)/N_c}\mh^{(N_F-\nl )/N_c}\,\,,
\eq
\bq
\Bigl (\cmb\Bigr )^2=\frac{1}{\mh}\Bigl (\la^{\bo}\det m\Bigr )^{1/N_c}=\la^{\bo/N_c}\,
\ml^{\nl/N_c}\,\mh^{(N_F-N_c-\nl )/N_c}=\frac{\ml}{\mh}\Bigl (\cma\Bigr)^2\,.
\eq

Clearly, at scales $\mu \gg \cma$ the large constituent masses $\mu_C^{{\it l}}=\cma$ and
$\mu_C^{{\rm h}}=\cmb$ are not yet formed, and all quarks behave as perturbative massless
particles. So, the fields $\Pi$ are not yet frozen, the factor $(\det \Pi)^{1/\nd}$ in (1)
is really $(\det {\ov Q}Q)^{1/\nd}$ and it is still a living
interaction, not a mass. As a result, there is still no difference between the (light at
lower scales) fields $\Pi$ and (heavy at lower scales) fields ${\ov Q} Q$. Therefore, the
projector $\cal P$ in curly brackets in (1) is still zero\,:
\bq
\cal P= \rm Tr \,({\ov Q}\Pi^{-1}Q)-N_F=0\,,\quad \mu>\cma>\cmb\,.
\eq

Now, the main point is that the projector $\cal P$ begins to be nonzero only after the
decreasing scale $\mu$ crosses the physical threshold at $\mu \sim \cmb < \cma$ (and not before,
at $\mu \sim \cma > \cmb$), where the mass gap between the heavy constituent quarks $(\oqh,\, \qh)^
{\rm (const)}$ with the masses $\cmb$ and the light pions $\Pi^{\rm h}_{\rm \ov h}=(\oqh\qh)
^{\rm (light)}$ with the masses $\sim \mh$\, appears and "begins to work", so that the fields
$\qh\,,\oqh$ become frozen. Before this, at $\mu> \cmb$, the constituent mass
$\cmb$ is not yet formed and the operator $\Pi^{\rm h}_{\rm \ov h}$ is not yet frozen and represents
rather two still living light quarks $\oqh \qh$, whose quantum
part still dominates over its c-number vacuum part. So, in the first term in $W_Q$ the
common factor $(\det \Pi)^{1/\nd}$ is not yet frozen completely, and still describes some
interaction, not a mass. Therefore, the constituent masses are not yet formed not only for
the $\oqh,\, \qh$ quarks, {\it but also for the $\oql,\, \ql$
quarks}. This shows that the very presence of  still living perturbative light
quarks $\oqh,\, \qh$ at $\cmb < \mu <\cma$ prevents also  the quarks $\oql,\, \ql$
from acquiring the large constituent mass $\cma$. So,
nothing happens yet at $\mu\sim \cma$  and the perturbative regime does not stop here, but
continues down to $\mu \sim\cmb$. This is the real physical threshold $\mu_H$, and the nonzero
non-perturbative contributions to the quark superpotential appear only after crossing this
region, and they appear simultaneously for all flavors.
\footnote{\,
In a sense, the constituent quarks can be thought of as the extended solitons.
And this shows also that the characteristic size of the heavier constituent quarks $\oql,\, \ql$
is not $R_{\it l}\sim 1/\cma$, but larger\,: $R_{\it l}\sim 1/\cmb \gg
1/\cma$, this is typical for a soft soliton, whose size is much larger than its
Compton   wavelength, $R^{(\rm soft)}_{\rm sol}\gg 1/M^{(\rm soft)}_{\rm sol}$. I.e., the size $R_{\it 
l}$ is the same as the size $R_{\rm h}$ of the lighter constituent quarks $\oqh,\, \qh\,:
R_{\rm h}\sim 1/\cmb $, which are, in this sense, the hard solitons.
}

So, at $\mu < \cmb$, instead of (1), the effective Lagrangian takes now the form\,:
\bq
L=\int \te \,\ote \, \Biggl \{ \rm {Tr}\,\sqrt {\Pi^\dagger \Pi}+Z_{\it l}\,{\rm Tr}_{\it l}\Biggl
( Q^\dagger e^{V} Q \Biggr )+Z_{\rm h}\,{\rm Tr}_{\rm h}\Biggl ( Q^\dagger e^{-V}
Q\Biggr )+\Bigl (Q\ra {\ov Q}\Bigr )\Biggr \}+\nonumber
\eq
\bq
+\int \te \,\Biggl \{-\frac{2\pi}{\alpha(\mu)}\,S\,+W_Q \Biggr \}\,,\quad
W_Q=\Biggl ( \frac{\rm {\det}\,\Pi}{\la^{\bo}}\Biggr )^{1/\ov N_c}\Biggl \{ \rm {Tr}
\Bigl (\ov Q \,\Pi^{-1}\, Q\Bigr )-N_F \Biggr \}+\rm {Tr} (m\,\Pi) \,,\nonumber
\eq
\bq
Z_l=\frac{m_l}{\cma}\,,\quad Z_{\rm h}=\frac{m_{\rm h}}{\cmb}\, .
\eq
Here\,: $\Pi$ is the total $N_F \times N_F$ matrix of all pions, and ${\ov Q},\, Q$ with ${\it l}$ or
${\rm h}$ - flavors are the constituent quarks with the masses $\cma$ or $\cmb$\,, respectively.

After integrating out all heavy constituent quarks (this leaves behind a large number of hadrons
made of constituent quarks which are weakly confined, the string
tension is $\sqrt \sigma\sim \lym\ll \cmb<\cma$) and proceeding in the same way as in
\cite{ch}, one obtains the same form as in \cite{ch}\,:\,\footnote{\,
It is only worth noting that because there is only one common threshold $\mu_H=
\cmb$ for all flavors, the renormalization factors $Z_{\pi}$ of all $N^2_F$ pions are the
same\,: $Z_{\pi}=z_Q^{-1}(\la,\,\cmb)\equiv z_Q^{-1}$, where $z_Q\ll 1$ is the perturbative
renormalization factor of the massless quark, see \cite{ch}.
}
\bq
L=\int \te \,\ote \, \Biggl \{ \rm {Tr}\,\sqrt {\Pi^\dagger \Pi}\Biggr \}\,\,
+\int \te \Biggl \{ -\frac{2\pi}{\alpha_{YM}(\mu, \Lambda_{L})}\, S
-N_F \Biggl ( \frac{\rm {\det}\,\Pi}{\la^{\bo}}\Biggr )^{1/\ov N_c}+
\rm {Tr}\Bigl (m \Pi \Bigr )\Biggr \}\,,\nonumber
\eq
\bq
 \Lambda_{L}^3=\Biggl ( \frac{\det\,\Pi}{\la^{\bo}}\Biggr )^{1/{\ov N_c}},\quad
  \langle\Lambda_L\rangle =\lym\ll \mu \ll \cmb\,.
\eq
So, the only difference with the case of the equal quark masses is that now the masses
entering $\rm Tr (m\Pi)$ are not equal.

Proceeding further as in \cite{ch} and going through the Veneziano-Yankielowicz (VY) procedure for
gluons \cite{VY}, one obtains that there is a large number of gluonia with masses $M_{\rm gl}\sim\lym
=(\la^{\bo}\det m)^{1/3N_c}$, and the lightest particles are the pions with the Lagrangian\,:
\bq
L_{\pi}=\int \te \,\ote \, \Biggl \{ \rm Tr \,\sqrt {\Pi^\dagger \Pi}\Biggr \}
+\int \te \Biggl \{ -\nd  \Biggl (\frac{\det\,\Pi}{\la^{\bo}}\Biggr )^{1/\ov N_c}+
\rm {Tr \Bigl ( m\, \Pi}\Bigr ) \Biggr \}\,,\,\,\, \mu\ll \Lambda_{YM}\,.
\eq
The pion masses are proportional to a sum of their two quark masses, i.e.\,:\, $M_{\pi}^{(ll)}=c_o
2m_l,\,\, M_{\pi}^{({\rm hh})}=c_o 2\mh,\,\, M_{\pi}^{(l{\rm h})}=c_o (\ml+\mh)$\,, where $c_o$ is a
constant $O(1)$.

Clearly, when the quark masses become equal, $\mh\ra \ml$, the Lagrangian (7) matches
smoothly those in (1) (and vice versa). This is as it should be, until both types of quarks
remain in the same $\rm DC$ - phase.

On the whole, it is seen that starting with the case of the equal quark masses and splitting
them smoothly, one obtains very similar results. The only essential restriction is that theory
has to stay in the $\rm DC_{\it l}-DC_{\rm h}$ phase. And the only new non-trivial point is that
there is only one common physical threshold $\mu_H$ where the non-perturbative effects turn on
and change the form of the Lagrangian, and this threshold is determined by the smallest diquark
condensate $\mu_H=\cmb<\cma$\,, see (2),(3).\\

Let us write finally the conditions for theory to be in the $\rm DC_{\it l}-DC_{\rm h}$ phase.
When going down from $\mu\sim \la$, the massless perturbative evolution will be stopped either
at $\mu_H=\cmb$ if $\cmb>\mh^{\rm pole}\,$, or at $\mu_H=\mh^{\rm pole}\,$ if $\mh^{\rm pole}>
\cmb$\,. So, the $\rm h$-quarks will be in the $\rm DC_{\rm h}$ phase at $\cmb>\mh^{\rm pole}$
and in the $\rm HQ_{\rm h}$ phase at $\cmb<\mh^{\rm pole}$. The phase transition occurs at
$\cmb \sim \mh^{\rm pole}$. Using (3), one obtains that the $\rm DC_{\rm h}$ phase persists
from $r=(\ml/\mh)=1$\, down to $r>r_1$\,:
\bq
\cmb=\mh^{\rm pole}=\mh\Bigl (\frac{\la}{\mh^{\rm pole}} \Bigr )^{\gamma_{+}}\,\ra\,
r=\frac{\ml}{\mh}=r_1\equiv \Bigl (\frac{\mh}{\la}\Bigr )^{\sigma}\ll 1,\, \sigma=
\frac{1}{\nl}\Bigl [\frac{2N_c}{1+\gamma_{+}}-(N_F-N_c)\Bigr ],\nonumber
\eq
\bq
3N_c/2<N_F<3N_c\,:\quad \gamma_{+}=\frac{\bo}{N_F}\,\quad\ra \quad \sigma=
\frac{\bo}{3 \nl}\,;\nonumber
\eq
\bq
N_c<N_F<3N_c/2\,:\quad \gamma_{+}=\frac{2N_c-N_F}{N_F-N_c}\quad \ra \quad
\sigma=\frac{N_F-N_c}{\nl}\,.
\eq
In (8)\,: $\gamma_{+}$ is the quark anomalous dimension. It is known in the conformal window
while, to have definite answers, the value $\gamma_{+}=(2N_c-N_F)/(N_F-N_c)$ used in \cite{ch}
for $N_c<N_F<3N_c/2$ is used also here and below in the text.

However, (8) is not the only condition, as if $r=\ml/\mh$ will be too small at $\nl<N_c$\,,
then $\cma$ will become larger than $\la$ and the $\it l$-quarks will be higgsed.
This happens, see (2), at\,:
\bq
\nl<N_c\,:\,\,\, \cma=\la\,\,\,\ra\,\,\, r=\frac{\ml}{\mh}=r_2\equiv \Bigl (\frac{\mh}{\la}
\Bigr )^{\frac{N_F-N_c}{N_c-\nl }}\ll 1\,.
\eq

On the whole, the theory is in the $\rm DC_{\it l}-DC_{\rm h}$ phase at\,:
\bq
a) \quad(\ml/\mh)>r_1 \quad{\rm for}\quad \nl>N_c\,;\quad b)\quad(\ml/\mh)>{\rm max}\,
(r_1,\,r_2)\quad {\rm for}\quad \nl<N_c\,;\nonumber
\eq
\bq
r_2>r_1\quad {\rm at}\quad \nl<N_o\,, \quad N_o=\left\{
\begin{array}{l l}
N_c\,\bo/2N_F  &   \quad{\rm for}\quad 3N_c/2<N_F<3N_c
\\
N_c/2 &   \quad{\rm for}\quad N_c<N_F<3N_c/2\quad.
\end{array}
\right.
\eq
\vspace{0.5cm}

{\bf 3\,. \,\, The \, {\boldmath  $\rm \quad DC_{\it l}-HQ_{\rm h}$} \bf phase }\,: $\quad \cma\ll
\mh^{\rm pole}\,,\quad\quad \nl>N_c$\,\,.\\

a) $\quad 3N_c/2<N_F<3N_c\,,\,\,3N_c/2<\nl<N_F$.\\

This is the different phase when the lighter $\it l$-quarks $\ql,\,\oql$ are in the
DC phase, while the heavier $\rm h$-quarks $\qh,\,\oqh$ are in the HQ phase.

For definiteness, let us agree to use below the following procedure. Theory is defined at
$\mu=\la$ by the values of quark masses\,: $\ml\equiv \ml(\mu=\la)\leq \mh\equiv \mh(\mu=\la)
\ll \la$. Starting with $\ml=\mh$, the unequal quark masses will be obtained with $\mh$
staying intact, while $\ml$ will become smaller, $\ml\ll\mh\ll \la$.

At $r\equiv \ml/\mh=1$ theory is in the $\rm DC_{\it l}-DC_{\rm h}$ phase, with the highest
physical scale $\mu_H$ given by $\mu_H=\cmb=\cma\ll \la$. As was explained in
section 2, the constituent masses of $\ql,\,\oql$ quarks can't be formed alone, but only after
all flavors will be frozen. So, as $r$ begins to decrease, the highest physical scale $\mu_H$
is determined by a competition between $\cmb<\cma$ and the pole mass $\mhp$ of $\qh,\,
\oqh$ quarks, $\mhp=\mh(\mu=\mhp)$.

The $\rm DC_{\it l}-DC_{\rm h}$ phase persists until $\cmb>\mhp$, while at $\cmb <\mhp$
the coherent condensate of $\qh, \oqh$ quarks can't be maintained any more, so that there is a
phase transition from the $\rm DC_{\it l}-DC_{\rm h}$ phase to the $\rm DC_{\it l}-HQ_{\rm h}$
one. This happens at $r\sim r_1\ll 1$\,, see (8)\,.

Although ar $r<r_1$ theory is in the $\rm DC_{\it l}-HQ_{\rm h}$ phase, there are two different
regimes (see section 4 below), depending on whether $r<r_1^{\prime}\ll r_1$, or $r_1^{\prime}
<r< r_1$, with $r_1^{\prime}$ determined by
\bq
\cma=\mhp \quad \ra \quad
r=r_1^{\prime}=\Biggl (\frac{\mh}{\la}\Biggr )^{\rho}\ll r_1\ll 1\,,\quad
\rho=\frac{1}{\nl-N_c}\Bigl [\frac{2N_c}{1+\gamma_{+}}-(N_F-N_c) \Bigr ]\,.
\eq

The regime at $r<r_1^{\prime}\ll r_1$ is much simpler and is considered first in this section.
So, let us take $r\ll r_1^{\prime}$ and consider the properties of this $\rm DC_{\it l}-HQ_{\rm h}$
phase. Here, the highest physical scale $\mu_H$ is given by the pole mass $\mhp=
\la(\mh/\la)^{1/(1+\gamma_{+})}\ll \la$ of the heavier quarks $\qh,\,\oqh$, see (8).

The condition $r\ll r_1^{\prime}$, i.e. $\cma\ll \mhp$, see (11), means that even if
$\ql,\,\oql$ quarks were trying to freeze in the threshold region around $\mu\sim \mhp$
by forming the largest possible constituent mass $\mu_C^{\it l}=\cma$, this is impossible if
$\cma\ll\mhp$, as even this mass will be too small for freezing.
Therefore, no non-perturbative effects turn on at $\mu \sim \mhp$ in this case, and the
region $\mu\sim \mhp$ is crossed in the pure perturbative regime.

At $\mu<\mhp$ the heavy $\rm h$-quarks decouple from the lower energy theory and can be integrated out.
What remains, is the lower energy theory with $N_c$ colors and $\nl>3N_c/2$ light flavors $\ql,\,
\oql$, which at $\mu_H^{\prime}<\mu\ll \mu_H=\mhp$ will be also in the conformal
regime ($\mu_H^{\prime}$ is the new highest physical scale of this lower energy theory). Let
us denote by ${\hat\Lambda}_Q$ the scale parameter of the new gauge coupling.
Its value can be found from the following considerations. At $\mhp
<\mu\ll\la$ the original coupling $\alpha(\mu)$ is already frozen at the value $\alpha
^{*}_1=O(1)$. At $\mu\ll\mhp$ the new coupling will be also frozen at a new value
$\alpha^{*}_2=O(1),\,\,\alpha^{*}_2>\alpha^{*}_1$. So, going from $\mu\ll \mhp$ up to $\mu
\sim\mhp$, the coupling of the lower energy theory becomes living in the interval $\delta
\mu\sim \mhp$ around $\mu=\mhp$, where it decreases significantly from
$\alpha^{*}_2$ to $\alpha^{*}_1$. This is only possible if the scale factor ${\hat
\Lambda}_Q$ of the lower energy theory is ${\hat\Lambda}_Q\sim \mhp$.

So, at $\mu<{\hat \Lambda}_Q$ we remain with $N_c$ colors, $\nl>3N_c/2$ light
flavors with the small current masses ${\hat m}_{\it l}\equiv \ml(\mu={\hat \Lambda}_Q)=z_Q^
{-1}(\la,\,\mhp)\,\ml\ll {\hat\Lambda}_Q$, and the coupling $\alpha(\mu)$ with the scale
parameter ${\hat\Lambda}_Q=\mhp$. Moreover, the value of the diquark condensate of {\it l} -flavors
is, see (3),(8)\,\,:
\bq
\langle (\oql\,\ql)_{\mu={\hat \Lambda}_Q}\rangle\equiv \delta^{\it l}_{\it \ov l }\,
\Bigl (\hcma \Bigr )^2\,,\quad \hcma=z_Q^{1/2}(\la,\,\mhp)\,\cma\ll {\hat \Lambda}_Q=\mhp\,,
\eq
\bq
z_Q(\la,\,\mhp)=\Bigl (\frac{\mhp}{\la}\Bigr )^{\gamma_{+}}=\Bigl (\frac{\mh}{\la}
\Bigr )^{\frac{\gamma_{+}}{1+\gamma_{+}}}\,\,\,, \quad \gamma_{+}^{\rm conf}=\bo/N_F\,.\nonumber
\eq

The properties of such a theory have been described in \cite{ch}\,-\, it is in the $DC_{\it l}$
phase. Its highest physical scale is $\mu_H^{\prime}=\hcma\ll{\hat\Lambda}_Q$, so that at $
\hcma<\mu\ll {\hat\Lambda}_Q$ it is in the conformal regime, while below the threshold at $\mu
\sim \hcma$ the quarks $\ql,\,\oql$ acquire the constituent masses $\mu_C^{\it l}=\hcma$ and
there appear $\nl^2$ light pions. The low energy Lagrangian of these pions at $\mu\ll \lym=
({\hat \Lambda}_{Q}^{3N_c-\nl }\det{\hml})^{1/3N_c}=(\la^{3N_c-N_F}\det m)^{1/3N_c}$ is\,:
\bq
L_{\pi}=\int \ote\te \,\sqrt{\rm Tr \,{\hat \Pi}^{\dagger}_{\it l}{\hat \Pi}_{\it l}}+\int \te
\Biggl \{-(\nl-N_c)\Biggl (\frac{\det \hat \Pi_{\it l}}{{\hat\Lambda}_Q^{3N_c-\nl }}\Biggr )^{1/
(\nl-N_c)}+{\hml}{\rm Tr \,{\hat \Pi_{\it l}}}\Biggr \}\,.
\eq

The normalization of the pion fields $\hat \Pi_{\it l}\equiv (\oql\,\ql)_{\mu={\hat \Lambda}_Q}
\,,\,\,\langle \hat \Pi_{\it l}\rangle=({\hat \Lambda}_Q^{3N_c-\nl }{\hat m}_{\it l}^{\nl-N_c})^
{1/N_c}$ is the most natural one, from the viewpoint of the lower energy theory. But it is also
useful to rewrite (13) with the "old normalization" of fields at $\mu=\la,\, \Pi_{\it l}
\equiv (\oql\ql)_{\mu=\la}\,,\,\, \langle \Pi_{\it l}\rangle=\langle S\rangle/\ml
\equiv\lym^3/\ml=\ml^{-1}(\la^{\bo}\det m)^{1/N_c}$. Then it looks as \,:
\bq
L_{\pi}=\int \ote\te \Biggl \{\,\,z_Q(\la,\,\mhp)\,\sqrt{\rm Tr \, \Pi^{\dagger}_
{\it l}\Pi_{\it l}}\,\,\Biggr \}+
\eq
\bq
+\int \te \Biggl \{-(\nl-N_c)\Biggl (\frac{\det \Pi_{\it l}}{\la^{\bo}\,\det \mh}\Biggr )^
{1/(\nl-N_c)}+{\ml}{\rm Tr \,{ \Pi_{\it l}}}\Biggr \}\,,\,\,\,
z_Q(\la,\mhp)=\Bigl (\frac{\mh}{\la}\Bigr )^{\bo/3N_c}\ll 1\,.\nonumber
\eq

On the whole, the mass spectrum includes in this case : a) a large number of heaviest
$\rm hh$-hadrons with their mass scale $\sim \mhp$\,,\,\, b) a large number of $\it {ll}$-
mesons with masses $\sim {\hat{\cal M}_{\rm ch}^{\it (l)}}$ made of non-relativistic quarks
$\ql,\,\oql$ with the constituent masses ${\hat{\cal M}_{\rm ch}^{\it (l)}}\ll \mhp$\,,\,\, c)
a large number of hybrid ${\rm h}{\it l}$-mesons made of above constituents (all quarks
are weakly confined, the string tension is $\sqrt \sigma\sim \lym\ll \hcma\ll\mhp$\,)\,,\,\, d)
a large number of gluonia with masses $\sim\lym\equiv (\la^{3N_c-N_F}\det m)^{1/3N_c}\ll{\hat{
\cal M}_{\rm ch}^{\it (l)}},\,\, \det m\equiv \ml^{\nl} \,\mh^{N_F-\nl }$,\,\, e) and finally,
$\nl^2$ lightest $\it l$\,-pions ${\hat\Pi}_{\it l}$ with masses $M^{\it l}_{\pi}
\sim {\hml}=z_Q^{-1}(\la,\,\mhp)\,\ml\ll\lym$.\\

b) $\quad 3N_c/2<N_F<3N_c,\,\,\, N_c<\nl<3N_c/2\,$.\\

The difference with the case "a" above is that at $\mu<\mu_H=\mhp$ and after the
heaviest quarks $\qh,\,\oqh$ have been integrated out, the lower energy theory is not in the
conformal regime at $\mu^{\prime}_H<\mu<\mhp$, but in the strong coupling
one, see \cite{ch}. I.e., its new coupling grows in a power-like fashion at $\mu\ll\mu_H=
\mhp$. This allows to determine its new scale parameter $\Lambda^{\prime}$, from
matching of couplings at $\mu=\mu_H=\mhp$, where both are $O(1)$. This is only
possible with $\Lambda^{\prime}=\mu_H=\mhp\equiv {\hat \Lambda}_Q$. Therefore, at
$\mu<{\hat \Lambda}_Q$ we remain with $N_c$ colors, $N_c<\nl<3N_c/2$ light
flavors with the current masses $\hml\equiv \ml(\mu={\hat \Lambda}_Q)\ll{\hat
\Lambda}_Q$, and the coupling with the scale parameter ${\hat \Lambda}_Q$. All this is
exactly as it was in the case "a" above, only the value of $N_{\it l}$ is smaller now.

As was explained in \cite{ch}, only the perturbative behavior in the interval of scales
$\mu^{\prime}_H=\hcma<\mu<{\hat \Lambda}_Q$ differs in this case from the conformal
behavior in the case "a" above, while at $\mu<\hcma$ all properties and mass spectra are
the same. In particular, the lowest energy pion Lagrangian is the same as in (13),(14), etc.\\

c) $\quad N_c<N_F<3N_c/2,\,\,\, N_c<\nl<N_F\,$.\\

In this case the original theory (at $\mu_H=\mhp<\mu<\la$) and the lower energy one (at $\mu<\mu_H)$
are both in the strong coupling regime. Their couplings $\alpha_{\pm}(\mu)$ will be matched at $\mu=
\mu_H=\mhp$, where $\mhp$ is the pole mass of $\qh,\, \oqh$ quarks: $\mhp=\mh(\mu=\mhp)$.
Because $\mhp \ll \la$, the upper (i.e. original) coupling $\alpha_{+}(\mu=\mhp)$
is parametrically large, and so will be $\alpha_{-}(\mu=\mhp)$.
Therefore, it is clear that its scale parameter $\Lambda^{\prime}\gg \mhp$.

To obtain definite expressions, let us make a (sufficiently weak) assumption that at $N_c<N_F
<3N_c/2$ the quark perturbative anomalous dimension $\gamma_Q$ is constant in the infrared
region. Then the quark renormalization factor $z^{+}_Q(\Lambda,\,\mu)$ and the coupling $a_{+}
\equiv N_c\alpha_{+}/2\pi$ of the higher energy theory, and  $z^{-}_Q(\Lambda^{\prime},\,\mu)\,
,\,\,a_{-}\equiv N_c\alpha_{-}/2\pi$ of the lower energy one behave as \cite{ch}\,:
\bq
z^{+}_Q(\la,\,\mu)=\Bigl (\frac{\mu}{\la}\Bigr )^{\gamma_{+}}\ll 1\,,\quad
a_{+}(\mu)=\Bigl (\frac{\la}{\mu}\Bigr )^{\nu_{+}}\gg 1\,,\,\,\, \nu_{+}=\frac{N_F\gamma_{+}-
\bo}{N_c}>0\,,\,\,\,\mu\ll \la\,, \nonumber
\eq
\bq
z^{-}_Q(\Lambda^{\prime},\,\mu)=\Bigl (\frac{\mu}{\Lambda^{\prime}}\Bigr )^{\gamma_{-}}\ll 1
\,,\quad a_{-}(\mu)=\Bigl (\frac{\Lambda^{\prime}}{\mu}\Bigr )^{\nu_{-}}\gg 1\,,\quad \nu_{-}
=\frac{\nl\gamma_{-}-\bo^{\prime}}{N_c}>\nu_{+}\,,\quad \mu\ll \Lambda^{\prime}\,,\nonumber
\eq
\bq
\bo=(3N_c-N_F),\,\,\, \bo^{\prime}=(3N_c-\nl )\,,
\eq
and the matching of couplings at $\mu=\mu_H=\mhp$ takes the form\,:
\bq
\Bigl (\frac{\la}{\mhp}\Bigr )^{\nu_{+}}=\Bigl (\frac{\Lambda^{\prime}}{\mhp}\Bigr )^{\nu_{-}}\,,
\quad \mhp=\frac{\mh}{z^{+}_Q(\la,\,\mhp)}=\mh\,\Bigl ( \frac{\la}{\mh} \Bigr )^
{ \frac{\gamma_{+}} {1+\gamma_{+}} }\gg \mh \,.
\eq
Because $\nu_{-}>\nu_{+}>0$\,, it is seen from (16) that $\mhp\ll \Lambda^{\prime}\ll\la$\,.
\footnote{\,
As a concrete example, one can use the values from \cite{ch}\,:
$\gamma_{+}=(2N_c-N_F)/(N_F-N_c)
\,,\,\,\gamma_{-}=(2N_c-\nl )/(\nl-N_c)\,,\,\, \nu_{+}=(3N_c-2N_F)/(N_F-N_c)\,,\,\,
\nu_{-}=(3N_c-2 \nl)/(\nl-N_c)\,$. The value of $\Delta$ in (18) is\, $0<\Delta=
(N_F-\nl )/(3N_c-2 \nl)<1/2$\, in this case.
}

Therefore, after the heaviest quarks $\qh,\, \oqh$ have been integrated out at $\mu<\mhp$\,, we
have now  $N_c$ colors, $N_c<\nl<3N_c/2$ flavors and the gauge coupling with the scale parameter
$\Lambda^{\prime},\,\, \mhp\ll \Lambda^{\prime}\ll\la$\,, determined from (16). The value of the
current mass $m_{\it l}^{\prime}$ of $\ql,\,\oql$ quarks and the pion fields $(\Pi^{\prime}_{\it
l})^{\ov l}\equiv({\oql}\ql)^{(\rm light)}_{\mu=\Lambda^{\prime}}$
normalized at $\mu=\Lambda^{\prime}$ look as\,:
\bq
m_{\it l}^{\prime}\equiv\ml(\mu=\Lambda^{\prime})= z^{-}_Q(\Lambda^\prime,\,\mhp)
\,{\hat m}_{\it l}\,,\,\, z^{-}_Q(\Lambda^\prime,\,\mhp)=\Bigl (\frac{\mhp}
{\Lambda^{\prime}}\Bigr )^{\gamma_{-}}\ll 1\,,\,\, {\hat m}_{\it l}=\ml\,
\Bigl ( \frac{\la}{\mh} \Bigr )^{\frac{\gamma_{+}}{1+\gamma_{+}}}\,,\nonumber
\eq
\bq
\langle (\Pi^{\prime}_{\it l})^i_{\ov j}\rangle \equiv \delta^i_{\ov j}\,\Pi^{\prime}_{\it l}
\,,\quad \Pi^{\prime}_{\it l}= \frac{1}{z^{-}_Q(\Lambda^\prime,\,\mhp)}\, {\hat \Pi}_
{\it l}\,,\quad {\hat \Pi}_{\it l}={\Pi}_{\it l}\, z^{+}_Q(\la,\,\mhp)=
{\Pi}_{\it l}\,\Bigl ( \frac{\mh}{\la} \Bigr )^{\frac{\gamma_{+}}{1+\gamma_{+}}}\,.
\eq

Therefore \cite{ch}, the low energy pion Lagrangian will have the form (13), with the
replacements\,: ${\hat \Pi}_{\it l}\ra \Pi^{\prime}_{\it l}\,,\,\,{\hat m}_{\it l}\ra
m_{\it l}^{\prime}\,,\,\,{\hat \Lambda}\ra \Lambda^{\prime}$. The pion mass is now $m_{
\it l}^{\prime}$. Being expressed through the pion fields ${\Pi}_{\it l}$  normalized at
$\mu=\la$, its superpotential has the universal form (14), and only the $Z^{\it l}_{\pi}$-factor
multiplying the Kahler term of $\it l$-pions is different, now it is:\,
\bq
Z^{\it l}_{\pi}=\Biggl (\frac{\mh}{\la}\Biggr )^{\Delta}\,,\quad \Delta={\frac{\delta}{1+\gamma_{+}}}
\,,\quad \delta=\Bigl [\gamma_{+}-\gamma_{-}\,\Bigl (\frac{\nu_{+}}{\nu_{-}}\Bigr )\Bigr ]
\,,\quad m_{\it l}^{\prime}=m_{\it l}/Z^{\it l}_{\pi}\,.
\eq

So, in the case considered, there are in the mass spectrum\,: a) the heaviest $\rm h$
-hadrons with their mass scale $\sim \mhp$ given by (16);\,
b) the $\it {ll}$-mesons made of the non-relativistic quarks $\ql,\, \oql$ with the
constituent masses $\mu_C^{\prime}=\langle\Pi^{\prime}_{\it l}\rangle^{1/2}$ (17);\, c) the
hybrid ${\rm h}{\it l}$-mesons made of the above constituents;\, d)
the gluonia with the universal mass scale $\lym$; d)\, and $\nl^2$
lightest $\it l$-pions with the masses $m_{\it l}^{\prime}\ll \lym$\,,\, see (18).\\

On the whole for this regime of the $\rm DC_{\it l}-HQ_{\rm h}$ phase and $\nl>N_c$\,,\, the
hierarchy of scales in the mass spectrum is always the same:\\
a) the largest masses are the pole masses $\mhp\ll \la$ of  $\qh,\, \oqh$ quarks ;\\
b) the next ones are the constituent masses $\mu_C^{\it l}$ of  $\ql,\,\oql$ quarks, they
are always much smaller than $\mhp$, but their concrete values
depend on the case considered\,\,;\\
c) the next one is the universal mass scale of gauge particles, it is always $\lym=(\la^
{\bo}\det\, m)^{1/3N_c}$;\\
d) the lightest are $\nl^2$  $\it l$-pions, their low energy Lagrangian has the universal
form (14), but the value of the $Z^{\it l}_{\pi}$-factor in front of the Kahler term (and so their
masses $M_{\pi}^{\it l}$\,) depends on the case considered.\\

{\bf 4\,.\,\,The {\boldmath  $\quad \rm DC_{\it l}-HQ_{\rm h}$} \bf phase }\,: $\quad \cmb\ll
\mhp\ll \cma$\,.\\

Let us consider now the most difficult regime with $r_1^{\prime}\ll r\ll r_1$\,, i.e. $\cmb
\ll \mhp \ll \cma$.

Let us trace the RG-flow when the running scale $\mu$ starts with $\mu=\la$ and decreases. As was argued
in section 2, even the large value of the running coherent condensate $\cma(\mu)$ does not mean, by
itself, that the large constituent mass $\cma(\mu)$ of $\ql,\,\oql$ quarks is really already formed,
as the projector $\cal P$ in (4) becomes  nonzero only after decreasing $\mu$ reaches such a value
$\mu_2$ that {\it both flavors, $\it l$ and $\it h$\,,} entering $\det({\ov Q}Q)$ acquire
masses larger than $\mu_2$ and become frozen. Therefore, the first point where this can happen
in the $\rm DC_{\it l}-HQ_{\rm h}$ phase with $\cmb\ll\mhp\ll\cma$ is the pole mass $\mhp$. So, there
is the narrow threshold region $\mu_2=\mhp/({\rm several})<\mu<\mu_1=({\rm several})\,\mhp$
around $\mhp$\,, where the non-perturbative effects turn on at $\mu_1$ and saturate at
$\mu_2$. In a sense, what will be going on in this transition region is qualitatively similar
to those described in section 2 for the $\rm DC_{\it l}-DC_{\rm h}$ phase, only the role
played by the coherent condensate $\cmb$ of the $\qh,\, \oqh$ quarks plays here their
perturbative pole mass $\mhp$. Therefore, all flavors become frozen in the threshold region
$\mu_2<\mu<\mu_1$ around $\mhp$. The $\qh,\,\oqh$ quarks - because their evolution
is stopped by their pole mass $\mhp$, while the $\ql,\, \oql$ quarks - because
their large constituent mass $\cma\gg \mhp$ is formed in this threshold region.

So, what form will the superpotential take at $\mu<\mhp$, after the non-perturbative
RG-flow has finished, and all quark masses become frozen\,? ( The heaviest are the constituent
$\ql,\,\oql$ quarks with masses $\cma$, the next ones are the $\qh,\, \oqh$ quarks with masses
$\mhp$, and the lightest are the pions $\Pi_{\it l}$ with the masses $\ml$\,, plus all
gluons which are still massless). Let us look at the superpotential in (1) or (5). Because
there are only $\Pi_{\it l}$ - pions, while the $\rm h$ - quarks are in the $\rm HQ$ phase
and there is no difference between $(\oqh \qh)$ and $\Pi_{\rm \ov h}^{\rm h}$, the $\rm h$-quark
contributions cancel in the projector $\cal P= \rm Tr \,({\ov Q}\Pi^{-1}Q)-N_F$, and it will
have the form\,: ${\cal P}={\rm Tr}\,(\oql \Pi^{-1}_{\it l}\ql )-\nl$\,. Now, what form can
$\det \Pi$ in (1) or (5) take at $\mu<\mhp$, after the evolution of all quark degrees of
freedom has been finished and the quarks $\qh,\,\oqh$ have been integrated out ?  In other
words, what their fields $\Pi^{\rm h}_{\rm \ov h}=(\oqh\qh)$ will be substituted by in $\det \Pi$ ?
The only possible form is\,:\,\footnotemark
\bq
\Pi^{\rm h}_{\rm \ov h}=\Bigl (\oqh \qh\Bigr )\quad\ra
\quad \Bigl (\,m^{-1}\,\Bigr )^{\it h}_{\rm \ov h}\,\,\Biggl (
\frac{\det \Pi_{\it l}}{\la^{\bo}\det m_{\it h}}\Biggr )^{1/(\nl-N_c)}\quad,
\eq
\bq
\Biggl (\frac{\det \Pi}{\la^{\bo}}\Biggr )^{1/(N_F-N_c)}\quad \ra \quad
\Biggl (\frac{\det \Pi_{\it l}}{\la^{\bo}\det \mh}\Biggr )^{1/(\nl-N_c)}\quad . \nonumber
\eq
\footnotetext{\,
The form (19) is determined uniquely by symmetries\,: a) the flavor symmetry
$SU(\nl)_L\times SU(\nl)_R\times SU(\nh )_L\times SU(\nh )_R$\,,\,\,
b) by the $R$-charges of the higher energy theory\,: $R(\ql)=R(\oql)=
R(\qh)=R(\oqh)=R(\Pi_{\it l})/2=(N_F-N_c)/N_F\,,\,\, R(\ml)=R(\mh)=2N_c/N_F$\,,\, c) by the
$R^{\prime}$-charges of the lower energy theory\,: $R^{\prime}(\ql)=R^{\prime}(\oql)=
R^{\prime}(\Pi_{\it l})/2=(\nl-N_c)/\nl\,,\,\, R^{\prime}(\qh)=R^{\prime}(\oqh)=1\,,\,\,
R^{\prime}(\ml)=2N_c/\nl\,,\,\,R^{\prime}(\mh)=0$\,, d) the overall normalization in (19) is
determined by the Konishi anomaly\,:\, $\mh \langle \,{\ov Q}^{\rm h}
Q_{\rm h}\rangle=\langle S\rangle$\,, see also (2).
}

So, instead of (5), the effective Lagrangian at $\mu<\mhp$ takes now the form
(let us recall that all fields entering (1),\,(5) and (20) are normalized at $\mu=\la$)\,:
\bq
L=\int \te \,\ote \, \Biggl \{ \rm {Tr}\,\sqrt {\Pi_{\it l}^\dagger \Pi_{\it l}}+
Z_{\it l}\,{\rm Tr}_{\it l}\Bigl ( Q^\dagger e^{V} Q\Bigr )+Z_{\rm h}{\rm Tr}_{\rm h}\Bigl
( Q^\dagger e^{V} Q\Bigr )+\Bigl ( Q\ra {\ov Q}\Bigr )\Biggr \}+\nonumber
\eq
\bq
+\int \te \,\Biggl \{ -\frac{2\pi}{\alpha(\mu)}\,S +W_Q \Biggr \} \,,\quad
Z_{\it l}=\frac{\ml}{\cma}\,,\quad Z_{\rm h}=\frac{\mh}{\mhp}\,\,,
\eq
\bq
\mhp=\frac{\mh}{z^{+}_Q(\la,\,\mhp)}=\mh\,\Bigl ( \frac{\la}{\mh}
\Bigr )^{ \frac{\gamma_{+}} {1+\gamma_{+}} }\gg \mh \,\,,\nonumber
\eq
\bq
W_Q=\Biggl ( \frac{\rm {\det}\,\Pi_{\it l}}{\la^{\bo}\det \mh}\Biggr )^{1/(\nl-N_c)}\Biggl \{ 
{\rm Tr}_{\it l} \Bigl ({\ov Q} \,\Pi_{\it l}^{-1}\, Q\Bigr )-\nl  \Biggr \}+\mh
{\rm Tr}_{\rm h}\Bigl ({\ov Q} Q \Bigr )+\ml \rm {Tr}\Bigl (\Pi_{\it l} \Bigr )\,\,.\nonumber
\eq

The meaning of (20) is the same as before in (1) or (5). All terms with the quark fields are
retained only to recall the values of their masses and, besides, it is implied that they can
be used, for instance, for some calculations where these quarks appear as valence ones. If
one is not interested in all this at $\mu<\mhp$, all quark terms in (20) can be omitted.

Now, let us write the explicit form of the inverse Wilsonian coupling \cite{SV}
$2\pi/\alpha_{W}(\mu)$
in (20). The simplest way to obtain it is to write out the result of the overall RG-flow from
$\mu=\la$ down to $\mu_2=\mhp/(\rm several)$ (because the RG is a group). So, one obtains\,:
\bq
\frac{2\pi}{\alpha_{W}(\mu_2)}=N_c\ln\Bigl (\frac{\mu_2^3}{\la^3} \Bigr )-\ln \Bigl (\frac{
\det \mu^{\it l}_{C}}{\la^{\nl} } \Bigr )-\nh\ln\Bigl (\frac{\mhp}{\la} \Bigr )+
\nl\ln\Bigl (\frac{1}{Z_{\it l}}\Bigr )+\nh\ln\Bigl (\frac{1}{Z_{\rm h}}\Bigr ).
\eq
In (21) the specific properties for the case considered are\,: a) the $Z_{\rm h}$-factor of
the $\qh,\,\oqh$ quarks is $Z_{\rm h}=\mh/\mhp$, because their mass $\mh(\mu)$ started
with the value $\mh$ at $\mu=\la$ and finished with the value $\mhp$ at $\mu=\mu_2$\,,\,\,
b) the constituent mass $\mu_C^{\it l}$ of the $\ql,\,\oql$ quarks in (21) has the form,
see (20)\,:
\bq
\Bigl (\mu^{\it l}_C\Bigr )^{\ov j}_i=\frac{1}{Z_{\it l}}\Biggl ( \frac{\rm \det \,\Pi_{\it
l}}{\la^{\bo}\det \mh}\Biggr )^{1/(\nl-N_c)}\Bigl (\Pi_{\it l}^{-1}\Bigr )^{\ov j}_i\quad.
\eq
So, the coupling in (20) at $\lym\ll\mu<\mu_2=\mhp/(\rm several)$ is weak and has the form (see also 
section 2 in\cite{ch}) \,:
\bq
\hspace{-5mm}\frac{2\pi}{\alpha_{W}\Bigl (\mu,\Lambda_L\Bigr )}=\frac{2\pi}{\alpha
\Bigl(\mu,\Lambda_L\Bigr )}-N_c\ln\frac{1}{{\rm g}^2(\mu,
\langle\Lambda_L\rangle)}=3N_c \ln\Bigl ( \frac{\mu}{\Lambda_L}\Bigr ),\nonumber
\eq
\bq
\Lambda^3_L=
\Biggl ( \frac{\det \Pi_{\it l}}{\la^{\bo}\det m_{\it h}}\Biggr )^{1/(\nl-N_c)}\,,\quad
\langle \Lambda_L\rangle=\Bigl (\la^{\bo}\det m \Bigr )^{1/3N_c}= \lym\,,
\eq
and the Lagrangian at $\mu<\mu_2$ looks as\,:
\bq
L=\int \te \,\ote \, \Biggl \{ \rm {Tr}\,\sqrt {\Pi_{\it l}^\dagger \Pi_{\it l}}\Biggr \}+
\eq
\bq
\int \te \,\Biggl \{ -\frac{2\pi}{\alpha(\mu,\Lambda_L)}\,S -
\nl\Biggl ( \frac{\rm {\det}\,\Pi_{\it l}}{\la^{\bo}\det \mh}\Biggr )^{1/(\nl-N_c)}+
\rm {Tr}\Bigl (\ml\Pi_{\it l} \Bigr )\,\Biggr \}\,.\nonumber
\eq
It describes gluonia with the universal  mass scale $M_{\rm gl}\sim \lym$ (interacting with
the pions $\Pi_{\it l}$), and after integrating them out through the VY-procedure
\cite{VY}, one obtains finally the lowest energy Lagrangian of pions\,:
\bq
L=\int \te \,\ote \, \Biggl \{ \rm {Tr}\,\sqrt {\Pi_{\it l}^\dagger \Pi_{\it l}}\Biggr\}+
\int \te \Biggl \{-({\it \nl}-N_c)\Biggl ( \frac{\rm {\det}\,\Pi_{\it l}}{\la^{\bo}\det \mh}
\Biggr )^{1/({\it \nl}-N_c)}+\rm {Tr}\Bigl (\ml\Pi_{\it l} \Bigr )\,\Biggr \}\,.
\eq
It describes $\nl^2$ $\it l$-pions $\Pi_{\it l}$ with the masses $\sim\ml$, and their
superpotential
has the standard universal form for the $\rm DC_{\it l}-HQ_{\rm h}$ phase, see (14).\\

 On the whole, the mass spectrum includes in this case\,: a) the $\it ll$-hadrons made of the
$\ql\,,\,\oql$\,-\, quarks with the constituent masses $\mu_C^{\it l}=\cma\ll \la$\,,
\,\, b) the $\rm hh$-hadrons made of the non-relativistic $\qh\,,\,\oqh$-quarks with the pole
masses $\mhp\ll\cma$\,,\,\,c) the hybrid ${\rm h}{\it l}$-hadrons made of
the above constituents (all quarks are weakly confined, the string tension is
$\sqrt \sigma\sim \lym\ll \mhp\ll \cma$)\,,\,\, d) the gluonia with their universal mass scale
$M_{\rm gl}\sim \lym\ll \mhp$\,,\,\, e) and finally, $\nl^2$ lightest $\it l$-pions
with the masses $\sim \ml\ll\lym$ and the superpotential (25) which is universal one for
the $\rm DC_{\it l}-HQ_{\rm h}$ phase . In a sense, this mass spectrum is similar to those
described in section 3, the main difference is that the hierarchy $\mhp\gg \mu_C^{\it l}$
from section 3 is reversed here.\\

Finally, let us look how the mass spectrum changes on both sides around the phase transition
at $r\sim r_1$, with $\cmb\sim \mhp \ll \cma$, see (8) and section 2.

a) The $\rm {h}$-flavors. In the $\rm DC_{\it l}-DC_{\rm h}$ phase at $r>r_1$\,, there are many
heavy $\rm {h}$-hadrons made of the non-relativistic $\qh\,,\,\oqh$-quarks with the
constituent masses $\cmb$\,, see (3), and $N_{\rm h}^2$ light $\rm {h}$- pions
with the masses $\sim \mh$. As $r$ crosses $r_1$, the coherent condensate of the
$\rm h$-flavors breaks down and theory appears in the $\rm DC_{\it l}-HQ_{\rm h}$ phase. The above
$N_{\rm h}^2$ light $\rm {h}$-pions with the masses $\sim \mh$ disappear from
the mass spectrum. At the same time, because $\cmb=\mhp$, the constituent masses
$\cmb$ of $\qh,\,\oqh$ quarks are substituted smoothly by their perturbative pole masses
$\mhp$, so that the mass spectrum of the heavy $\rm {h}$-hadrons made now
of the non-relativistic current quarks $\qh\,,\,\oqh$ changes smoothly.

b) The $\it {l}$-flavors. In the $\rm DC_{\it l}-DC_{\rm h}$ phase at $r>r_1$\,, there are many
heavy $\it {l}$-hadrons made of the $\ql\,,\,\oql$ - quarks with the
constituent masses  $\cma\gg\cmb$\,, see (2), and $\nl^2$ lightest $\it{l}$-pions with the
masses $\sim \ml\ll \mh$. In the $\rm DC_{\it l}-HQ_{\rm h}$ phase with $\cma\gg \mhp$ at $r<r_1$\,,
all these $\it l$-hadrons and $\nl^2$ \, $\it l$-pions are still present in the
spectrum and their masses remain the same.

c) The hybrid ${\rm h}{\it l}$-flavors. In the $\rm DC_{\it l}-DC_{\rm h}$ phase at $r>r_1$\,,
there are many heavy ${\rm h}{\it l}$-mesons with the masses $(\cmb+\cma)$ and
corresponding ${\rm h}{\it l}$-pions with the small masses $(\mh+\ml)$. In the $\rm DC_
{\it l}-HQ_{\rm h}$ phase at $r<r_1$\,, these light hybrid pions are absent. As for the heavy hybrid
mesons, their masses change smoothly from $(\cmb+\cma)$ to $(\mhp+ \cma)$.

d) Finally, all gluons remain massless down to the scale $\mu\sim \lym$, and there is a large
number of gluonia with the same masses $M_{\rm gl}\sim \lym$ in both phases.\\

{\bf 5\,.\,\, The \,{\boldmath $\rm Higgs_{\it l}-DC_{\rm h}$} and
{\boldmath $\rm Higgs_{\it l}-HQ_{\rm h}$}\,\,\bf phases}.\quad
{\boldmath $\cma>\la\,,\,\,\,\nl<N_c-1$}\,.\\

There are only two different phases at $\nl>N_c$ (because always $\cma\ll \la$, and the lighter
quarks are never higgsed): a) $\rm DC_{\it l}-DC_{\rm h}$, and b) $\rm DC_{\it l}-HQ_{\rm h}$.

At  $\nl<N_c$\,, in addition to the above two phases, two new phases appear at $\cma>\la$,
when the lighter $\it l$-quarks are higgsed, $\langle\ql\rangle=\langle\,\oql\rangle\neq 0$
\,, while the heavier quarks are either in the DC phase, or in the HQ phase.

So, let us take $r\ll r_2$ ( see (9), but not too small, see below) and look for a mass
spectrum in this phase. One can
proceed in a close analogy with the case of the Higgs phase for $N_F<N_c-1$ in \cite{ch}, the
only difference is that not all flavors are higgsed now, only $\ql,\,\oql$.

So, one begins with the scale of the large gluon mass, \,$\mu=\mu_{\rm gl}={\rm g}_H{\hcma}
\gg \la\,,\,\, {\rm g}^2_H=4\pi\alpha(\mu=\mu_{\rm gl})\ll 1\,,\,\, \langle Q^{\it l}_{a}
\rangle_{\mu=\mu_{\rm gl}}=\delta^{\it l}_{a}\,\hcma,\,\,\,\langle {\ov Q}_{\it \ov l}^{a}
\rangle_{\mu=\mu_{\rm gl}}=\delta_{\it \ov l}^{a}\,\hcma\,$. The gauge symmetry $SU(N_c)$ is broken
down to $SU(N_c-\nl)$ at this high scale $\mu_H=\mu_{\rm gl}$\, and $(2\nl N_c -\nl ^2)$ gluons
become massive. The same number of degrees of freedom of $\ql,\,\oql$ quarks acquire
the same masses and become the superpartners of these massive gluons (in a sense, they can be
considered as the heavy "constituent quarks"), and there remain $\nl^2$  light complex pion fields
${\hat\Pi}_{\it l}=({\ov Q}_{\it \ov l}Q^{\it l})_{\mu=\mu_{\rm gl}
}\,,\,\,\langle {\hat \Pi}_{\it l}\rangle=\,(\hcma)^2$. The
value of $\ml(\mu)$ at this scale is $\hml\equiv \ml(\mu=\mu_{\rm gl})$ (this will be the
$\it l$-pion mass), and similarly, the mass of h-quarks at this scale is $\hmh\equiv
\mh(\mu=\mu_{\rm gl})$. Besides, let us denote as ${\hat \Pi}_{\rm hl}$ and ${\hat \Pi}_{\rm lh}$ the
hybrids (in essense, these are the h-quark fields $Q^{\rm h}_a,\, {\ov Q}_{\rm \ov h}^{a}$ with broken
colors $a=1...\nl$)\,, while $\qh\,,\,\oqh$ will be now the h-quark fields with unbroken colors.

Consider first the case $\nl<\bo/2$, i.e. $b^\prime_o=(\bo-2\nl)>0$. After integrating out all
heaviest particles with masses $\sim \mu_{\rm gl}$ and proceeding in the same way as in \cite
{ch}, one obtains the lower energy Lagrangian at the scale $\mu\lesssim\mu_{\rm gl}$\,:
\bq
L=\int\te\,\ote\,\Biggl\{2\,\rm {Tr}\,\sqrt {{\hat\Pi_{\it l}}^{\dagger} {\hat\Pi_{\it l}}}
+{\rm Tr}_{\rm h}\,\Biggl ( {\hat Q}^{\dagger} e^{\hat V} {\hat Q}+ \Bigl (\hat Q\ra \hat{\ov Q}\Bigr )
\Biggr )+{\rm Tr}\,\Bigl ({\hat\Pi}_{\rm hl}^{\dagger}{\hat\Pi}_{\rm hl}+ {\hat\Pi}_{\rm lh}
^{\dagger}{\hat\Pi}_{\rm lh} \Bigr )+\cdots\Biggr \}\,+ \nonumber
\eq
\bq
+\int \te \Biggl \{-\frac{2\pi}{\alpha(\mu, \hat\Lambda)}\hat S+ \hml \rm {Tr}\, {\hat \Pi_{\it
l}}+\hmh \rm {Tr}_{\rm h}\,\Bigl (\hat{\ov Q} \hat Q\Bigr ) +\hmh \rm {Tr} \Bigl ({\hat \Pi}_{\rm lh}
{\hat \Pi}_{\rm hl}\Bigr ) \Biggr \}\,,\nonumber
\eq
\bq
{\hat\Lambda}^{b^\prime_o}=\frac{\la^{\bo}}{z_Q^{\nl} \det {\hat\Pi_{\it l}}}\,\,\Bigl
(\frac{z_Q^\prime}{z_Q}\Bigr )^{N_F-\nl }\,,\quad b^\prime_o=(\bo-2 \nl)>0\,,
\eq
\bq
z_Q=z_Q(\mu_{\rm gl},\,\la\,|\,N_c,\,N_F)=\frac{\hml}{\ml}\,,\quad
z_Q^\prime=z_Q(\mu_{\rm gl},\,\langle\hat\Lambda\rangle\,|\,N_c-\nl ,\,N_F-\nl )=
\frac{\hmh}{\mh^\prime}\,.
\eq

Here\,:\, $\hat S={\hat W}^2_{\alpha}/32\pi^2\,,\,\, {\hat W}_{\alpha}$ are the gauge field strengths
of $(N_c-\nl)^2-1$ remaining massless gluon fields, $\alpha(\mu, \hat\Lambda)$ is the gauge coupling
of this lower energy theory and $\hat\Lambda$ is its scale parameter, $z_Q\ll 1$ is the massless
quark renormalization factor from $\mu=\mu_{\rm gl}$ down to $\mu=\la$ in the original theory
with $N_c$ colors and $N_F$ flavors, $z_Q^\prime\ll 1$ is the analogous renormalization factor
from $\mu=\mu_{\rm gl}$ down to $\mu=\langle\hat\Lambda\rangle$ in the lower energy theory
with $N_c-\nl$ colors and $\nh$ remained active h-flavors $\qh\,,\,\oqh$\,,\,\, $\mh^\prime\ll
\langle\hat \Lambda\rangle$ is the current mass of $\qh\,,\,\oqh$-quarks in this lower energy
theory at $\mu=\langle\hat\Lambda\rangle$.
\footnote{\,
Both $z_Q$ and $z^\prime_Q$ are only logarithmic in the case considered.
}
All fields in (26) are normalized at $\mu=\mu_{\rm gl}$.
Finally, the dots in (26) denote residual D-term interactions. It is supposed that these play no
significant role for the case considered in this section and will be neglected in what follows.

Therefore, the hybrids ${\hat\Pi}_{\rm hl}$ and ${\hat\Pi}_{\rm lh}$ will appear in the
spectrum as (weakly interacting) particles with the masses $\hmh$.
These hybrids will not be written explicitly below (but implied).

The lower energy theory with $N_c^\prime=N_c-\nl$ colors, $N_F^\prime=\nh$ flavors of $\qh,\,
\oqh$-quarks, with $b^\prime_o>0$  and $\mh^\prime\ll \langle\hat\Lambda\rangle$ will be in
the $DC_{\rm h}$ - phase \cite{ch}. The constituent mass $\mu_C^{\rm h}=(z_Q^\prime)^{1/2}
\hcmb\ll \langle\hat\Lambda\rangle$ is formed in the threshold region $\mu\sim \mu_C^{\rm
h}$ and there appear $\nh^2$ \, $\rm h$-pions $\Pi^\prime_{\rm h}\,,\,\,\langle
(\Pi^\prime_{\rm h})^i_{\ov j}\rangle=\delta^i_{\ov j}\,(\mu_C^{\rm h})^2$\,,\, with
masses $\mh^\prime$. After integrating out these constituent h-quarks, one remains with the
Yang-Mills theory with $N_c^\prime=N_c-\nl$ colors and with the new scale factor $\Lambda_L$
of the gauge coupling, and with $\nh^2$ \, h-pions \cite{ch}\,:
\bq
L=\int \te \,\ote \, \Biggl \{2\, \rm {Tr}\,\sqrt {{\hat \Pi_{\it l}}^\dagger {\hat\Pi_{\it l}
}}+\rm {Tr}\,\sqrt { (\Pi^{\prime}_{\rm h})^\dagger \Pi^{\prime}_{\rm h}}\Biggr \}\,
\eq
\bq
+\int \te \Biggl \{ -\frac{2\pi}{\alpha(\mu, \Lambda_L)}\,{\hat S}
-N^\prime_F\Biggl (\frac{\det \Pi^{\prime}_{\rm h}}{{\hat\Lambda}^{b^\prime_o}}
\Biggr )^{1/{(N^\prime_F-N^\prime_c)}}+ \hml \rm {Tr}\, {\hat \Pi_{\it l}}+{\mh^\prime}
\rm {Tr}\,{\Pi^{\prime}_{\rm h}}\Biggr \}\,,\nonumber
\eq
\bq
\Lambda_L^3=\Biggl (\frac{\det \Pi^{\prime}_{\rm h}}{{\hat\Lambda}^{b^\prime_o}}
\Biggr )^{1/{(N^\prime_F-N^\prime_c)}}\,,\quad \langle \Lambda_L\rangle=\lym =\Bigl (\la^{\bo}
\det m \Bigr )^{1/3N_c}\,,\quad \det m=\ml^{\nl} \,\mh^{N_F-\nl }\,.\nonumber
\eq

Proceeding through the VY-procedure, one obtains the lowest energy pion Lagrangian\,:
\bq
L=\int \te \,\ote \, \Biggl \{2\, \rm {Tr}\,\sqrt {{\hat \Pi_{\it l}}^\dagger {\hat\Pi_{\it l}
}}+\rm {Tr}\,\sqrt { (\Pi^{\prime}_{\rm h})^\dagger \Pi^{\prime}_{\rm h}}\Biggr \}\,+\nonumber
\eq
\bq
+\int \te \Biggl \{
-(N^\prime_F-N^\prime_c)\Biggl (\frac{\det \Pi^{\prime}_{\rm h}}{{\hat\Lambda}^{b^\prime_o}}
\Biggr )^{1/{(N^\prime_F-N^\prime_c)}}+ \hml \rm {Tr}\, {\hat \Pi_{\it l}}+
{\mh^\prime}\rm {Tr}\,{\Pi^{\prime}_{\rm h}}\Biggr \}\,.
\eq
Substituting $\hat\Lambda$ from (26), this takes the form\,:
\bq
L=\int \te \,\ote \, \Biggl \{2\, \rm {Tr}\,\sqrt {{\hat \Pi_{\it l}}^\dagger {\hat\Pi_{\it l}
}}+\rm {Tr}\,\sqrt { (\Pi^{\prime}_{\rm h})^\dagger \Pi^{\prime}_{\rm h}}\Biggr \}\,+\nonumber
\eq
\bq
+\int \te \Biggl \{-(N_F-N_c)\Biggl (\frac{{\rm z}^{N_F}_Q\det {\hat\Pi}_{\it l}\det \Pi^
{\prime}_{\rm h}}{\la^{\bo}\,({\rm z}^{\prime}_Q)^{N_F-\nl }}
\Biggr )^{1/{(N_F-N_c)}}+ \hml \rm {Tr}\, {\hat \Pi_{\it l}}+
{\mh^\prime}\rm {Tr}\,{\Pi^{\prime}_{\rm h}}\Biggr \}\,.
\eq

The Lagrangian (30) (with the hybrid pions ${\hat \Pi}_{\rm hl}$ and ${\hat \Pi}_{\rm lh}$ reinstated),
being expressed through the fields $\Pi_{\it l},\,\Pi_{\rm h},\, \Pi_{\rm hl},\, \Pi_{\rm lh}$ and the
masses $\ml\,,\,\mh$ normalized at the "old scale" $\mu=\la$\,, takes the form\,:
\bq
L=\int \te \,\ote \, \Biggl \{\frac{2}{\rm z_Q}\,\rm {Tr}\,\sqrt { \Pi_{\it l}^\dagger
\Pi_{\it l}}+\frac{z^\prime_Q}{z_Q}\,\rm {Tr}\,\sqrt { \Pi_{\rm h}^\dagger
\Pi_{\rm h}}+\,\frac{1}{z_Q}\, \rm {Tr}\Bigl (
{\Pi}_{\rm hl}^{\dagger}{\Pi}_{\rm hl}+ {\Pi}_{\rm lh}^{\dagger}{\Pi}_{\rm lh}
\Bigr )+\cdots\Biggr \}\,+\nonumber
\eq
\bq
+\int \te \Biggl \{-(N_F-N_c)\Biggl (\frac{\det \Pi_{\it l}\det \Pi_{\rm h}}
{\la^{\bo}}\Biggr )^{1/(N_F-N_c)}+ \ml\,{\rm Tr}\,  \Pi_{\it l}+
\mh\,{\rm Tr}\,\Bigl (\Pi_{\rm h}+{\Pi}_{\rm hl}{\Pi}_{\rm lh}\Bigr )\Biggr \}\,,\nonumber
\eq
\bq
\rm \Pi^{\prime}_{\rm h}=\frac{z^\prime_Q}{z_Q}\,\Pi_{\rm h}\,,\quad \hat \Pi_{\it l}=\frac{1}
{z_Q}\,\Pi_{\it l}\,,\quad \ml=\frac{\hml}{z_Q}\,,\quad \mh^\prime=\frac{\hmh}
{z^\prime_Q}=\frac{z_Q}{z^\prime_Q}\,\mh\,.
\eq

On the whole for this case when theory is deeply in the $\rm Higgs_{\it l}-DC_{\rm h}$ phase
(i.e. $\hcma\gg\la$), the mass spectrum looks as follows.
There is\,: a) $(2\nl N_c-\nl ^2)$  massive gluons and the same
number of their superpartners\, - \,the "constituent $\it l$-quarks" with heaviest masses
$\mu_{\rm gl}\gg\la$\,,\,\, b) a large number of hadrons made of non-relativistic
constituent $\qh,\,\oqh$-quarks with masses $\sim \mu_C^{\rm h}\ll \langle\hat\Lambda
\rangle\ll\la\ll\mu_{\rm gl}$\,,\,\, c) a large number of strongly coupled gluonia with the
mass scale $M_{\rm gl}\sim \lym \ll \mu_C^{\rm h}$\,,\,\,\, d) $\nh^2$\,\,\,  h-pions with masses
$\sim \mh^\prime\ll \lym$\,,\,\,e) the hybrid pions ${\Pi}_{\rm hl}$ and ${\Pi}_{\rm lh}$ (these are
$\qh$ and $\oqh$ - quarks with higgsed colors) with masses
$\hmh \ll\mh^\prime$\,,\,\,f)  $\nl^2$ \,\, lightest $\it l$-pions with masses $\hml\ll\hmh$.\\

At $\nl<N_c-1$\,, starting with $r\equiv \ml/\mh= 1$ when all quarks are in the DC phase and
diminishing $r$, there always will be a number of phase transitions. The $\rm DC_{\it l}-
DC_{\rm h}$ phase is maintained until (10) is fulfilled.

Let us take $\nl<N_o$\,, see (10). Then, as $r$ approaches $r_2$ from above, $\cma$ approaches
$\la$ from below, with all quarks in the $\rm DC_{\it l}-DC_{\rm h}$ phase.
When $\cma$ overshoots $\la$, there is a phase transition as the $\it l$-quarks become
higgsed. The crucial parameter here (i.e. at $\cma>\la$\,, but not too large, see below) is
$b^\prime_o=(3N^\prime_c-N^\prime_F)=(\bo-2\nl)$.\,\,\,The $\qh\,,\,\oqh$-quarks will be in the
$DC_{\rm h}$ phase at $b^\prime_o>0$, and in the $\rm HQ_{\rm h}$ phase at $b^\prime_o<0$.
If  $\nl<N_o$, then $\nl< \bo/2$ also, so that as $\cma$ overshoots $\la$ and the $\it
l$-quarks are higgsed, the $\rm DC_{\rm h}$-phase of  $\rm h$-quarks is maintained.

Let us trace how the mass spectrum changes on both sides around this phase transition between
the $\rm DC_{\it l}-DC_{\rm h}$ and $\rm Higgs_{\it l}-DC_{\rm h}$ phases, at $r\sim r_2\ll 1$
(see section 2).

a) The gluons. In the $\rm DC_{\it l}-DC_{\rm h}$ phase  at $r<r_2$\,, all $(N_c^2-1)$ gluons
can be thought of as having the small masses $M_{\rm gl}\sim \lym$. In the $\rm Higgs_{\it l}-
DC_{\rm h}$ phase at $r>r_2$\,, the $SU(N_c)$ gauge symmetry is broken down to the non-Abelian
$SU(N_c-\nl)$ one\,,\,\,$\nl<N_c-1$. So, $(2\nl N_c-\nl^2)$ gluons acquire the large masses $M_
{\rm gl}\sim \la\gg\lym$, while $(N_c-\nl)^2-1$ gluons remain with the same small
masses $\sim \lym$.

b) The $\it ll$-flavors. In the $\rm DC_{\it l}-DC_{\rm h}$ phase at $r<r_2$\,, the confined
$\ql\,,\,\oql$ quarks have large constituent masses $\mu_C^{\it l}=\cma\sim \la$ and there
are $\nl^2$ light $\it ll$-pions with small masses $M_{\pi}^{\it l}\sim\ml$. In the $\rm Higgs_
{\it l}-DC_{\rm h}$ phase at $r>r_2$\,, there is $(2\nl N_c-\nl^2)$  massive
quarks which are the superpartners of massive gluons and so have the same masses $\sim
\la$. In a sense, these quarks can be thought of as remnants of the previous constituent
$\it l$-quarks and their masses match smoothly across $r\sim r_2$\,. As for the $\it ll
$-pions, their number and masses also match smoothly across the phase transition.

c) The $\rm hh$-flavors. Nothing happens also with the confined constituent $\qh\,,\,\oqh$
-quarks (i.e. those with unbroken colors) with the masses $\mu_C^{\rm h}=\cmb\ll \la$,
and with $\nh^2$ \, $\rm hh$-pions with masses $\sim\mh\gg \ml$. But in the $\rm Higgs_{\it l}-DC_
{\rm h}$ phase at $r>r_2$\,, the $Q^{\rm h}\,,\,{\ov Q}_{\rm \ov h}$ - quarks with broken
colors appear now individually in the spectrum as light particles with the masses
$\sim (\mh-\ml)\simeq \mh$. They can be thought of as remnants of the previous hybrid
$\Pi_{\rm hl}$ and $\Pi_{\rm lh}$ - pions with the masses $(\mh+\ml)\simeq \mh$\,, which were present
in the spectrum in the $\rm DC_{\it l}-DC_{\rm h}$ phase at $r<r_2$.\\

Let us take now $\nl>N_o$. The theory is in the $\rm DC_{\it l}-DC_{\rm h}$ phase at $r=1$\,.
As $r$ decreases, there is first a phase transition to the $\rm DC_{\it l}-HQ_{\rm h}$ phase
at $r\sim r_1\gg r_2$, which persists until $r$ approaches $r_2$ from above. If $\nl>\bo/2$\,,
as $\cma$ overshoots $\la$ and the $\it l$ - quarks are higgsed, the $\rm HQ_{\rm h}$-phase of
the $\qh\,,\,\oqh$-quarks is maintained.

But there are such values of $N_c<N_F<3N_c$ and $\nl<N_c-1$ that $N_o<\nl<\bo/2$. In this case,
the theory stays in the $\rm DC_{\it l}-HQ_{\rm h}$ phase as $\cma$ approaches $\la$ from below,
while as $\cma$ overshoots $\la$ the $\rm h$-quarks condense and there appear $\Pi_{\rm h}$
-pions, and theory will be in the $\rm Higgs_{\it l}-DC_{\rm h}$ phase. So, not only
the $\it l$-quarks change their phase, but the $\qh\,,\,\oqh$-quarks also. The reason for
this is the following. At $\cma$ slightly above $\la$ when the $\it l$-quarks are already higgsed,
the remaining lower energy theory has $\langle\hat\Lambda\rangle\sim \la\,,\,\,N^\prime_c=
N_c-\nl $ colors, $N^\prime_F=N_F-\nl$ flavors, and with $\mh\ll \la$ and $\cmb$ staying intact
because $\langle\hat\Lambda\rangle\sim \la$. But the pole mass ${\ov m}_{\it h}^{\rm pole}$ of
the $\qh\,,\,\oqh$-quarks is smaller now in this new theory than it was before higgsing,
${\ov m}_{\it h}^{\rm pole}\ll \mhp$, because the quark anomalous dimension diminished.
So, while the hierarchy was $\mhp\gg\cmb$ before higgsing, it is reversed now after higgsing,
${\ov m}_{\it h}^{\rm pole}\ll \cmb$, and the $\qh\,,\,
\oqh$-quarks also change their phase simultaneously with the $\ql\,,\,\oql$ ones.\\

This is not the end of the story with $b^\prime_o>0$ however, because to stay in the $\rm
Higgs_{\it l}-DC_{\rm h}$ phase the condition $\mh^\prime=\mh(\mu=\langle\hat\Lambda\rangle)
\ll \langle\hat\Lambda\rangle$ is necessary, so that $r=\ml/\mh$ has not to be too small. As
$r$ decreases at $\nl<N_c-1$ and $b^\prime_o>0$\,, $\langle\hat\Lambda\rangle$ in (26) decreases
in a power-like fashion because $\cma$ grows $\sim(1/r)^\omega\,,\,\,\,\omega=(N_c-\nl)/2N_c$\,,
see (2), ($\hcma\sim \cma$ up to a logarithmic factor), while $\mh^\prime$ changes only
logarithmically. So, as decreasing $r$ crosses the smaller value $r_3\ll r_2$ where decreasing
$\langle\hat\Lambda\rangle$ becomes $\langle\hat\Lambda\rangle <\mh^\prime$, there is the
phase transition from the $\rm Higgs_{\it l}-DC_{\rm h}$ phase to the $\rm Higgs_{\it l}-HQ_
{\rm h}$ one. The coherent condensate of the $\qh,\,\oqh$-quarks breaks down, the $\Pi
_{\rm h}$-pions disappear, and at $r\ll r_3$ the heavy $\qh,\,\oqh$-quarks will be in
the perturbative weak coupling regime, like $\rm h$-quarks with higgsed colors
(but weakly confined, the string tension is small\,: $\sqrt \sigma\sim \lym$)\,. I.e., the
lower energy theory at $\mu\ll\mu_{\rm gl}$ contains\,: the unbroken non-Abelian gauge group
$SU(N^{\prime}_c)\,,\,N^{\prime}_c=N_c-\nl$\,, with the scale factor $\hat\Lambda$ (26) of its
gauge coupling, and $N^{\prime}_F=N_F-\nl\,\,(\,N^{\prime}_c<N^{\prime}_F <3N^{\prime}_c$\,)\,
flavors of the heavy non-relativistic quarks $\qh\,,\,\oqh$ with their pole masses $\mh^{\rm pole}
\gg\langle\hat\Lambda\rangle$,\, plus the $\it l$-pions entering $\hat\Lambda$, see (26), and plus the
hybrid pions $\Pi_{\rm hl},\, \Pi_{\rm lh}$ (these are the light $\qh,\, \oqh$
quarks with higgsed colors,
weakly interacting through residual D-terms interactions). So, we don't give here further
detail, because this is a simple regime and it is evident how to deal with this case.
The mass spectrum in this $\rm Higgs_{\it l}-HQ_{\rm h}$ phase at $r\ll r_3$ looks as
follows.  There is\,: a) $(2\nl N_c-\nl^2)$ of massive gluons and the same number of their
superpartners\,-\,"the constituent $\it l$-quarks" with the heaviest masses $\mu_{\rm gl}\gg
\la$\,,\,\, b) a large number of hadrons made of non-relativistic $\qh,\,\oqh$-quarks
with the perturbative pole masses $\mh^{\rm pole}$\, (the hierarchies look here as\,:
$\langle\hat\Lambda\rangle\ll\lym\ll \mh^{\rm pole}\ll \la\,$, with $\hat\Lambda$ from (26),
while $\lym$ is the gauge coupling scale arising after the $\qh\,,\,\oqh$-quarks were
integrated out)\,,\,\,\, c) the hybrids $\Pi_{\rm hl},\, \Pi_{\rm lh}$ ( these are $\qh,\, \oqh$
quarks with higgsed colors) with the masses $\lym\ll\hat\mh=\mh(\mu=\mu_{\rm gl})\ll \mh^{\rm pole}$\,,
\,\,\, d) a large number of strongly coupled gluonia with the mass scale $M_{\rm gl}\sim \lym $\,,\,
\,e)\, $\nl^2$ lightest $\it l$-pions with the masses $2\hat\ml=2\ml(\mu=\mu_{\rm gl})=2 z_Q\,\ml$.
The lowest energy Lagrangian of these $\it l$-pions will have the same Kahler term as in (31),
and the same universal superpotential as in (14).\\

Finally, let us consider the case $\cma> \la\,,\,\,\nl<N_c-1\,,\,\,b^\prime_o=(\bo-2 \nl)<0$. As
was pointed out above, as $\cma$ approaches $\la$ from below, the theory is already in the
$\rm DC_{\it l}-HQ_{\rm h}$ phase, and as $\cma$ overshoots $\la$ and the $\it l$-quarks
become higgsed, the confined $\qh\,,\,\oqh$-quarks remain in the $\rm HQ_{\rm h}$ phase.
So, on the whole, this is the $\rm Higgs_{\it l}-HQ_{\rm h}$ phase. But now, at $b^\prime_o=
(\bo-2 \nl)<0$ and $\cma\gg \la\,,\,\,\langle \hat \Lambda\rangle \gg \cma$, see (26), and in
the interval of scales $\lym\ll\mu\ll\mu_{\rm gl}$ the remained non-Abelian $SU(N_c-\nl)$ gauge
theory with $N_F-\nl$ of  confined $\qh\,,\,\oqh$ quarks is in the weak coupling logarithmic
regime. On the whole, this is also a very simple case (see sections 2 and 8 in \cite{ch}), and
it is clear what will be the mass spectrum. Qualitatively, it is similar to those described in
the preceding paragraph with $b^\prime_o>0$ and in the same $\rm Higgs_
{\it l}-HQ_{\rm h}$ phase at $r<r_3$ (and the lowest energy Lagrangian of the lightest
$\it l$-pions will be the same), so that we are not going into further detail with this case.\\

{\bf 6\,.\,\,\,Conclusions\,.}\\

As was described above within the dynamical scenario considered in this paper, ${\cal N}=1$ SQCD
with $N_c$ colors (with the scale factor $\la$ of
their gauge coupling), $N_c<N_F<3N_c$ flavors of light quarks, with $N_{\it l}$ lighter flavors with
masses $\ml$ and $\nh=N_F-\nl$ heavier ones with masses $\mh$\,,\,\,$0<\ml<\mh\ll \la$\,, will be
in the different phase states, depending on the values of the above parameters. Besides, the
mass spectra are also highly sensitive to the values of these parameters.

The lighter $\ql\,,\oql$ -quarks may be in two different phases\,: either in the $\rm DC$
(diquark condensate) phase at $\cma\ll \la$ (both at $\nl<N_c$ and $\nl>N_c$), or in the $\rm
Higgs$ phase at $\cma\gg \la$ (at $\nl<N_c$ only). The heavier $\qh\,,\oqh$ -quarks may be
also in two different phases\,: either in the $\rm DC$ phase at $\cmb\gg\mh^{\rm pole}$, or in
the $\rm HQ$ (heavy quark) phase at $\cmb\ll \mh^{\rm pole}$. So on the whole, four different
phases are realized in this theory. For each of them, the mass spectra and the corresponding
interaction Lagrangians were described above in the text.
\footnote{\,
The $N_c$\,-\,dependence of various quantities (e.g. $\langle S\rangle\sim N_c^0$\,, etc.)
used everywhere above in the text is the same as in the main text in \cite{ch}. And as in
\cite{ch}, the correct $N_c$ dependence ($\langle S\rangle\sim N_c$\,, etc.) can be easily
reinstated (see section 9 in \cite{ch}).
}

We did not consider in this paper the Seiberg dual theories \cite{S}\cite{IS} with
unequal quark masses. As was argued in \cite{ch}, the direct and dual theories are not
equivalent even in a simpler case of equal quark masses. There are no chances that
the situation will be better for unequal quark masses.\\

This work is supported in part by the RFBR grant 07-02-00361-a.\\

\end{document}